\documentclass[runningheads]{llncs}
\usepackage{graphicx}
\usepackage{tikz}
\usepackage{amsmath}
\usepackage{hyperref}
\usepackage{multirow}
\usetikzlibrary{positioning}
\usetikzlibrary{decorations.text}
\usetikzlibrary{decorations.pathmorphing}
\usetikzlibrary{arrows.meta}
\usepackage[linesnumbered,ruled,vlined]{algorithm2e}

\SetCommentSty{mycommfont}

\SetKwInput{KwInput}{Input}                
\SetKwInput{KwOutput}{Output}              

\usetikzlibrary{shapes.callouts}
\tikzset{
  level/.style   = { ultra thick, blue },
  connect/.style = { dashed, red,-> },
  notice/.style  = { draw, rectangle callout, callout relative pointer={#1} },
  label/.style   = { text width=2cm }
}
\tikzset{%
  >={Latex[width=2mm,length=2mm]},
            base/.style = {rectangle, rounded corners, draw=black,
                           minimum width=4cm, minimum height=1cm,
                           text centered, font=\sffamily},
  activityStarts/.style = {base, fill=blue!30},
       startstop/.style = {base, fill=red!30},
    activityRuns/.style = {base, fill=green!30},
         process/.style = {base, minimum width=2.5cm, fill=orange!15,
                           font=\ttfamily},
}
\usetikzlibrary{arrows,automata,chains,matrix,positioning,scopes}
\makeatletter
\tikzset{join/.code=\tikzset{after node path={%
\ifx\tikzchainprevious\pgfutil@empty\else(\tikzchainprevious)%
edge[every join]#1(\tikzchaincurrent)\fi}}}
\makeatother
\tikzset{>=stealth',every on chain/.append style={join},
         every join/.style={->}}
\tikzstyle{labeled}=[execute at begin node=$\scriptstyle,
   execute at end node=$]

\begin{document}
\title{Test4Enforcers: Test Case Generation for Software Enforcers}
%
%
\author{Michell Guzman\inst{1} \and
Oliviero Riganelli\inst{1}\orcidID{0000-0003-2120-2894}\and
Daniela Micucci\inst{1}\orcidID{0000-0003-1261-2234}\and
Leonardo Mariani\inst{1}\orcidID{0000-0001-9527-7042}}
\authorrunning{Guzman et al.}
%
\institute{University of Milano-Bicocca, Milan 20126, Italy\\
\email{\{michell.guzman,oliviero.riganelli,daniela.micucci,leonardo.mariani\}@unimib.it}}
\maketitle              
\begin{abstract}
Software enforcers can be used to modify the runtime behavior of software applications to guarantee that relevant correctness policies are satisfied. Indeed, the implementation of software enforcers can be tricky, due to the heterogeneity of the situations that they must be able to handle. Assessing their ability to steer the behavior of the target system without introducing any side effect is an important challenge to fully trust the resulting system. To address this challenge, this paper presents Test4Enforcers, the first approach to derive thorough test suites that can validate the impact of enforcers on a target system. The paper also shows how to implement the Test4Enforcers approach in the DroidBot test generator to validate enforcers for Android apps.  

\end{abstract}

\section{Introduction}\label{sec:introduction}
To prevent undesired behaviors that may result in software failures and crashes, runtime enforcement techniques have been used to modify the runtime behavior of software systems, forcing the systems to satisfy a set of correctness policies~\cite{falcone2011runtime,ligatti2010theory}. So far, runtime enforcement has been already applied to multiple domains, including security~\cite{Khoury:SecurityEnforcement:ACMTISS:2012}, resource management~\cite{riganelli2017policy,Riganelli:ProactiveLibraries:ACMTAAS:2019}, and mobile computing~\cite{Falcone:AndoridEnforcement:RV:2012,Xu:Aurasium:Security:2012}.  

The enforcement logic is often defined with an input/output~\cite{Lynch:IO:PN:1988} or an edit automaton~\cite{Ligatti:Enforcements:TISS:2009} that is used to guide the implementation of the actual software enforcer. Although sometimes part of the code of the enforcer can be obtained automatically, the implementation of the final component that can be injected in the actual system normally requires the manual intervention of the developers, to make it fit well with the complexity of the runtime context. In particular, manual intervention is necessary to handle those aspects that are abstracted (i.e., not represented) in the enforcement model, such as handling the values of the parameters that are not represented in the model but must be used in function calls, adding the code to extract data from the runtime events produced by the monitored system, adding the code necessary to obtain security permissions, and more in general handling any other aspect or optimization not fully represented in the model.

As a consequence, the correctness of the resulting enforcer is threaten by three possible sources of problems: 
\begin{itemize}
\item \emph{Model inaccuracies}: This is the case of a wrong enforcement model that consequently leads to the implementation of a wrong software enforcer. Wrong models are the results of a bad design activity. When all the elements of the system can be accurately modelled (e.g., the environment, the monitored system, and the enforcer), this problem can be mitigated using verification techniques~\cite{riganelli2017verifying};
\item \emph{Inconsistent implementations}: This is the case of a software implementation that is not perfectly compliant with the model. This may happen when developers unintentionally introduce bugs
while implementing an enforcer starting from its enforcement model. In some cases, code generation techniques can be used to obtain part of the code automatically and mitigate this problem~\cite{Falcone:AndoridEnforcement:RV:2012};
\item \emph{Faulty additional code}: This is the case of a fault in the code that must be added to the software enforcer, which is distinct from the code that directly derives from the enforcement model, to obtain a fully operational enforcer that can be deployed in the target environment. The amount of additional code that is needed can be significant, depending on the complexity of the involved elements (e.g., the environment, the monitoring technology, and the monitored application). No simple strategy to mitigate the problem of verifying the correctness of this code is normally available.
\end{itemize}

To guarantee the correctness of a software enforcer before it is deployed in the target system, in addition to validating the enforcement model, it is necessary to extensively test the enforcer against all these possible threats. Compared to a regular testing scenario, testing software enforcers requires approaches that are able to target the specific events relevant to the enforcer and assess their impact on the runtime behavior of the system. In particular, \emph{test case generation should check that a software enforcer both modifies executions according to the expected strategy and leaves executions unaltered when the observed computation does not require any intervention}. 

In this paper we present Test4Enforcers, the first test case generation approach designed to address the challenge of testing the correctness of software enforcers. Note that enforcers typically act on the internal events of a software system, while test case generation works on its interface, and thus the testing strategy must discover \emph{how to activate the proper sequences of internal events from the application interface}. Test4Enforcers originally combines the knowledge of the enforcement strategy with GUI exploration techniques, to discover how to generate the right sequences of interactions that validate the interactions between the target software system and the  enforcer. The resulting test cases are executed on both the application with and without the enforcer in place to detect undesired behavioral differences (e.g., side effects) and unexpected similarities (e.g., lack of effect of the enforcer) to the tester. We also concretely show how to implement Test4Enforcers for Android apps by extending the DroidBot test case generation technique~\cite{Droidbot}. 

The paper is organized as follows. Section~\ref{sec:enforcer} provides background information about policy enforcement. Section~\ref{sec:hsi-method} presents the Test4Enforcers test case generation strategy. Section~\ref{sec:caseStudy} reports our experience with Test4Enforcers applied to an Android app. Section~\ref{sec:related} discusses related work. Finally, Section~\ref{sec:conclusions} provides concluding remarks.

\section{Policy enforcement}\label{sec:enforcer}
In this section we introduce the notion of runtime policy and policy enforcement. 


\subsection{Runtime Policy}
A runtime policy is a predicate over a set of executions. More formally, let $\Sigma$ be a finite set of observable program actions $a$. An \emph{execution} $\sigma$ is a finite or infinite non-empty sequence of actions $a_1;a_2;\ldots;a_n$. $\Sigma^*$ is the set of all finite sequences, $\Sigma^\omega$ is the set of infinite sequences, and $\Sigma^\infty = \Sigma^* \cup \Sigma^\omega$ is the set of all sequences. Given a set of executions $\chi \subseteq \Sigma^\infty$, a \emph{policy} is a predicate $P$ on $\chi$. A policy $P$ is satisfied by a set of executions $\chi$ if and only if $P(\chi)$ evaluates to $true$.

Policies may concern different aspects of the runtime behavior of a software, such as resource usages, security, and privacy.  For example, a policy about resource usage for the Android framework requires that anytime an app stops using the Camera, the app explicitly releases the Camera to make it available to the other applications~\cite{CameraAPI}. More precisely, \emph{''if an activity\footnote{Activities are fundamental components of Android apps and they represent the entry point for a user's interaction with the app \href{https://developer.android.com/guide/components/activities}{https://developer.android.com/guide/components/activities}} is using the camera and the activity receives an invocation to the callback method \texttt{onPause()}\footnote{\texttt{onPause()} is a callback method that is invoked by the Android framework when an activity is paused.}, the activity must release the camera}''. 
We use this policy throughout the paper to describe our approach.

\subsection{Policy Enforcement Models}

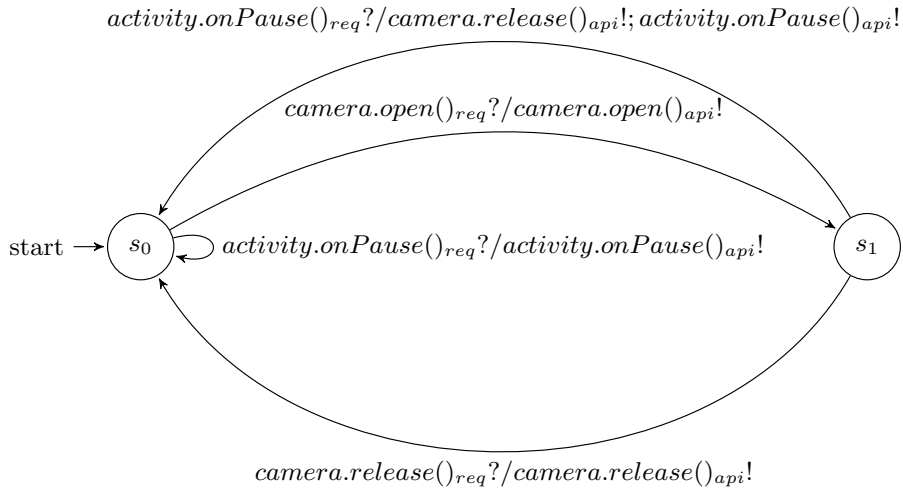
\begin{figure}
    \centering
    \resizebox{\textwidth}{!}{%
\begin{tikzpicture}[->,>=stealth',shorten >=1pt,auto,node distance=8.9cm,scale=1,transform shape]

  \node[state,initial] at (-4, 0) (a0p0e0) {$s_0$};

  \node[state] (e2p1a0) [right of=a0p0e0] {$s_1$};

 \path (a0p0e0) edge      [bend left]        node {$camera.open()_{req}?/camera.open()_{api}!$} (e2p1a0)

 (a0p0e0) edge      [loop right = 60]        node {$activity.onPause()_{req}?/activity.onPause()_{api}!$} (a0p0e0)

 (e2p1a0) edge      [bend left=60]        node {$camera.release()_{req}?/camera.release()_{api}!$} (a0p0e0)
 (e2p1a0) edge      [bend right = 60]        node[pos=0.5,above,rotate=0] {$activity.onPause()_{req}?/camera.release()_{api}!;activity.onPause()_{api}!$} (a0p0e0);
\end{tikzpicture}
}%
    \caption{Enforcer for systematically releasing the camera when the activity is paused.}
    \label{fig:enforcer-composition}
\end{figure}

A policy enforcement model is a model that specifies how an execution can be altered to make it comply with a given policy. Policy enforcers can be represented with both edit and input/output automata. In Figure~\ref{fig:enforcer-composition} we show the model of an enforcer specified as an input/output automaton that addresses the before-mentioned policy about the Camera. The inputs are requests intercepted at runtime (these events are labeled with the \emph{req} subscript) by the software enforcer and the outputs are the events emitted by the enforcer in response to the intercepted requests (these events are labeled with the \emph{api} subscript). When the label of the output is the same than the label of the input (regardless of the subscript), the enforcer is just forwarding the requests without altering the execution. If the output is different from the input, the enforcer is manipulating the execution suppressing and/or adding requests.  

When the current state is state $s_0$, the Camera has not been acquired yet and the \texttt{activity.onPause()} callback can be executed without restrictions. If the camera is acquired by executing \texttt{camera.open()} (transition from $s_0$ to $s_1$), the camera must be released before the activity is paused (as done in the transition with the input \texttt{activity.release()} from $s_1$ to $s_0$). If the activity is paused before the camera is released, the enforcer modifies the execution emitting the sequence \texttt{camera.release()} \texttt{activity.onPause()}, which guarantees that the policy is satisfied despite the app is not respecting it.

This enforcement strategy must be translated into a suitable software component to achieve runtime enforcement. With reference to the main classes of errors introduced in the paper, the following types of bugs may affects its implementation:
\begin{itemize}
\item \emph{Model Inaccuracies:} Although the designer may believe the enforcement model is correct, the strategy might end up being incorrect. For instance, the enforcer in Figure~\ref{fig:enforcer-composition} releases the camera when an activity is paused but does not acquire the camera back when the execution of the activity is resumed. This is a source of problems when the activity does not automatically re-acquire the camera once resumed. The strategy is thus inaccurate and should be extended to also include this stage. For simplicity in this paper we use the enforcer in Figure~\ref{fig:enforcer-composition} without complicating the model with the part necessary to acquire again a forcefully released camera.
\item \emph{Inconsistent Implementations:} The model must be translated into working code. Concerning the behavior specified in the model, the  corresponding code can be easily produced automatically, thus preventing inconsistencies (unless the generator is faulty). If the code is implemented manually, still the conformance of the code with the model has to be verified.
\item \emph{Faulty Additional Code:} In order to achieve code that can be deployed, a non-trivial amount of scaffolding code must be implemented, as well as many details should be worked out at the code level. For instance, the software enforcer derived from the model in Figure~\ref{fig:enforcer-composition} must be integrated with the monitoring solution used to capture events. This may already require a significant amount of code to be implemented. Moreover, although not shown in the model, an implementation of the enforcer that also acquires back the camera has to track both all the parameters used to initialize the camera and all the changes performed to these parameters to later recreate a correctly configured camera. These details are typically worked out by the engineers at the code level and are not present in the model.     
\end{itemize}
 
Test4Enforcers automatically generates tests that target the above listed issues.

\section{Test4Enforcers}\label{sec:hsi-method}

\begin{figure}
    \centering
        \includegraphics[width=\textwidth]{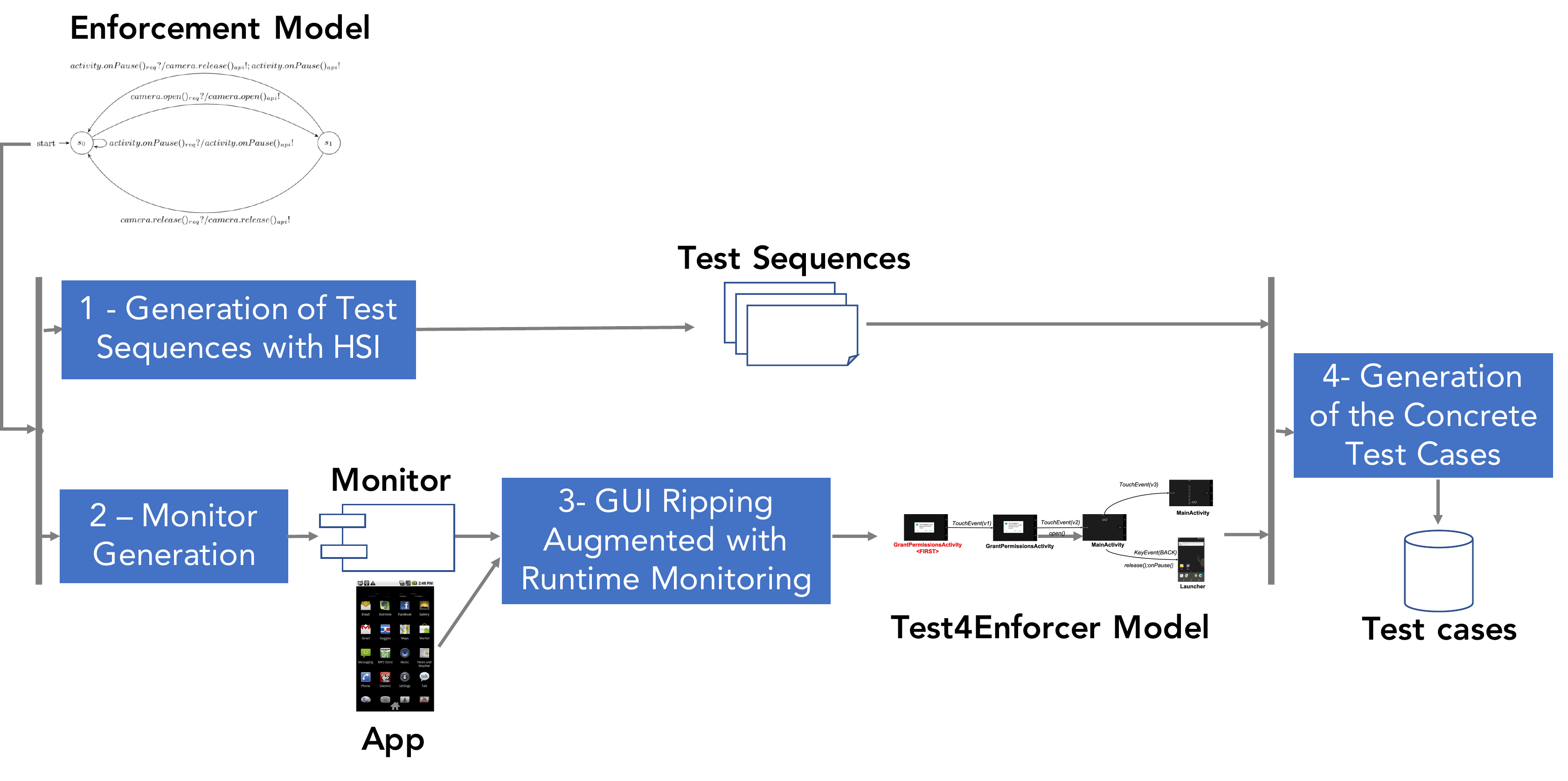}
    \caption{Test case generation with Test4Enforcers.}
    \label{fig:approach}
\end{figure}

Test4Enforcers generates test cases in 4 steps, as illustrated in Figure~\ref{fig:approach}. The first step, namely \emph{Generation of Test Sequences with HSI}, generates the test sequences that must be covered to thoroughly test the behavior of the enforcer based on the behavior specified in the enforcement model. To obtain concrete test cases that cover these sequences, Test4Enforcers explores (second and third steps) the application under test to determine what user interface (UI) interactions generate the events that belong to the alphabet of the enforcement model. In particular, the second step, namely \emph{Monitor Generation}, uses the events in the alphabet of the enforcer to obtain a monitor that observes when these events are executed. The third step, namely \emph{GUI Ripping Augmented with Runtime Monitoring}, runs a GUI Ripping process~\cite{Memon:Ripping:WCRE:2013} that systematically explores the UI of the application under test while logging the events relevant to the enforcer with the generated monitor. The output is the Test4Enforcers model, which is a finite state model that represents the GUI states that have been exercised, the UI interactions that cause transitions between these states, and the events relevant to the monitor that have been produced during transitions. Note that although we present the steps in this order, the second and third steps can be executed in parallel with the first step. Finally, the fourth step, namely \emph{Generation of the Concrete Test Cases}, looks for the sequences of UI interactions that exactly cover the test sequences identified in the first step and deemed as relevant to verify the behavior of the enforcer. These UI interactions, enriched with program oracles, are the test cases that can be executed to validate the activity of the enforcer. In the following, we describe each step in details.

\subsection{Generation of Test Sequences} \label{sec:hsi}
In this step, Test4Enforcers generates the sequences of operations that must be covered with testing based on the behavior specified in the enforcement model. A notable method to generate these sequences from finite state models is the \emph{W-method}~\cite{chow1978testing,lee1996principles,sidhu1989formal}. 

The \emph{W-method} has been designed to unveil possible errors in the implementation, such as, erroneous next-states, extra/missing states, etc. The main idea of the method is that starting from a state-based representation of a system, it is possible to generate sequences of inputs to reach every state in the model, cover all the transitions from every state, and identify all destination states to ensure that their counter-parts in the implementation are correct.

One limitation of the W-method is that it requires the model of the system to be \emph{completely specified}, that is, in every state of the model there must be a transition for every possible input. This is however not the case for enforcers, since the inputs are generated by other components and frameworks, and not all the combinations of inputs can be feasibly generated (e.g., \texttt{onPause()} and \texttt{onResume()} are callbacks produced when an activity is paused and resumed, respectively, as a consequence it is impossible to produce a sequence of two \texttt{onPause()} events without an intermediate \texttt{onResume()} event). In order to tackle this limitation, we use the \emph{Harmonized State Identifiers} (HSI) method, which is a variante of the W-method that does not require the model of the system to be completely specified~\cite{luo1995selecting,belli2015fault}. HSI exploits a few key concepts that are introduced below.

We first define an \emph{Input/Output Automaton} $A$ as a tuple $\langle S, s_0, sig, T \rangle$, where $S$ is a finite set of states; $s_0 \in S$ is the initial state; $sig$ is the set of actions of $A$ partitioned into input actions $in$, internal actions $int$, and output actions $out$; $T \subseteq S \times \{in \cup int\} \times out^*  \times S$ is a set of transitions (the symbol $^*$ denotes a sequence of actions). 
Note that differently from classic Input/Output Automaton, here we consider models that are not input-enabled, that is, every input cannot be received from every state, as mentioned due to the requirements of the environment where the enforcer is executed. Moreover, the automaton can produce zero or more outputs, denoted with $out^*$, in response to a single input as a consequence of the activity of the enforcer.

A sequence $in_1, \ldots, in_k$ with $in_j \in in, \forall j=1 \ldots k$ is an input sequence for state $s \in S$, if there exist states $s_1, \ldots, s_{k+1}$ such that $s=s_1$ and $\langle s_i, in_i, o, s_{i+1} \rangle \in T, \forall i=1 \dots k$ (note $o$ is not bounded to any value). $\Omega(s)$ is used to denote all input sequences defined for state $s$.
Similarly, given a state $s$ and an input sequence $\gamma=\langle in_1, \ldots, in_k \rangle \in \Omega(s)$, the function $\lambda(s, \gamma)$ denotes the output sequence $o_1, \ldots, o_m$ emitted by the automaton when accepting $\gamma$.

To generate an effective test suite, it is important to be able to distinguish the covered states. We say that two states $s_i,s_j \in S$ are \emph{distinguishable} if there exists a \emph{separating sequence} $\gamma \in \Omega(s_i) \cap \Omega(s_j),$ such that $\lambda ( s_i, \gamma)  \neq \lambda (s_j, \gamma)$, otherwise they are not distinguishable. 


To generate a thorough test suite, HSI exploits the notions of \textit{transition cover} and \textit{separating families}. We say that the set $P$ is a \textit{transition cover} of $A$ if for each transition $x$ from state $s$ there exists the sequence $\alpha x \in P$ such that $\alpha \in \Omega(s_0)$ and $s$ is the state reached by accepting $\alpha$. 
By definition, $P$ also includes $\epsilon.$\\

For instance, a transition cover $P$ for the enforcer in Figure~\ref{fig:enforcer-composition} is given by the following set\\


\noindent $P=\{$

\noindent \quad $\epsilon$, 

\noindent \quad $activity.onPause()_{req}$,

\noindent \quad $camera.open()_{req}$, 

\noindent \quad $camera.open()_{req}\ activity.onPause()_{req}$,

\noindent \quad $camera.open()_{req}\ camera.release()_{req}$

\noindent $\}$.\\



The sequences in the transition cover, each one defined to cover a different transition, are extended with actions aimed at determining if the right state has been finally reached once the sequence is executed. To this end, HSI computes the \emph{separating families}, which are sets of input sequences, one set for each state, that can be executed to distinguish a state from the other states of the system. In particular, a separating family is a set of input sequences $H_i \subseteq \Omega(s_i)$ for $s_i \in S$ satisfying the following condition: For any two distinguishable states $s_i, s_j$ there exist sequences $\beta \in H_i, \gamma \in H_j, $ such that $\alpha$ is a common prefix of $\beta$ and $\gamma$ and $\lambda(s_i, \alpha) \neq \lambda(s_j, \alpha).$

Computing the separating families for the automaton in Figure~\ref{fig:enforcer-composition} is straightforward since the two states have a single input in common that produces different outputs, allowing to distinguish the states. Thus\linebreak $H_0=H_1= \{ activity.onPause()_{req}\}$.

The HSI method can take into consideration the case the actual states of the implementation differ from the number of states in the model. However, we expect the software enforcer to have exactly the same number of states reported in the model. In such a case, the resulting test sequences are obtained by concatenating the transition coverage set $P$ with the separating families $H_i$. Note that the concatenation considers the state reached at the end of each sequence in $P$, namely $s_i$, to concatenate such a sequence with the sequences in the corresponding separating family, namely $H_i$.\\ 

In our example, this process generates the following sequences to be covered with test cases:\\


\noindent \quad $activity.onPause()_{req}$, 

\noindent \quad $activity.onPause()_{req}\ activity.onPause()_{req}$,

\noindent \quad $camera.open()_{req}\ activity.onPause()_{req}$, 

\noindent \quad $camera.open()_{req}\ activity.onPause()_{req}\ activity.onPause()_{req}$, 

\noindent \quad $camera.open()_{req}\ camera.release()_{req}\ activity.onPause()_{req}$ \\

HSI also includes a step to remove duplicates and prefixes from the generated set. Test4Enforcers only removes duplicates. In fact, removing a sequence that is a prefix of another sequence may drop a feasible test sequence in favour of an infeasible one (e.g., the longer sequence might be impossible to generate due to constraints in the environment, while it might be still possible to test the shorter sequence). In our example, since the list includes no duplicates, it is also the set of sequences that Test4Enforcer aims to cover to assess the correctness of the enforcer in Figure~\ref{fig:enforcer-composition}.

\subsection{Monitor Generation}

This step is quite simple. It consists of the generation of a monitor that can trace the execution of the events that appear in the enforcement model. Since we focus on enforcers that intercept and produce method calls, the captured events are either API calls (e.g., the invocation of \texttt{open()} and \texttt{release()} on the Camera) or callbacks (e.g., the invocation of \texttt{onPause()} on the activity). In this phase, the user of Test4Enforcer can also specialize the general enforcement strategy to the target application, if needed. For instance, the user can specify the name of the target activity that must be monitored replacing the placeholder name \texttt{activity} that occurs in the model with the name of an actual activity in the application (e.g., \texttt{MainActivity}). Multiple copies of the same enforcer can be generated, if multiple activities must be monitored. 

In our implementation, we consider Xposed~\cite{Xposed2020} as instrumentation library for Android apps. The monitoring module is implemented once and simply configured every time with the set of methods in the alphabet of the enforcer.

\subsection{GUI Ripping Augmented with Monitoring}

GUI Ripping is an exploration strategy that can be used to explore the GUI of an application under test with the purpose of building a state-based representation of its behavior~\cite{Memon:Ripping:WCRE:2013}. In particular, GUI ripping generates the state-based model of the application under test by systematically executing every possible action on every state encountered during the exploration, until a given time or action budget expires. Our implementation of Test4Enforcers targets Android apps and uses DroidBot~\cite{Droidbot} configured to execute actions in a breadth-first manner to build the state-based model. 

The model represents each state of the app according to its set of visible views and their properties. More rigorously, a state of the app under test is a set of views $s=\{v_i | i=1 \ldots n\} \in S_{app}$, and each view is a set of properties $v_i=\{p_{i1},\ldots, p_{ik}\}$. For instance, an \texttt{EditText} is an Android view that allows the users to enter some text in the app. The \texttt{EditText} has a number of properties, such as \texttt{clickable}, which indicates if the view reacts to click events, and \texttt{text}, which represents the text present in the view. 

Operations that change the set of visible views (e.g., because an activity is closed and another one is opened) or the properties of the views (e.g., because some text is entered in an input field) change the state of the app. DroidBot uses the following set of actions $A_{app}$ during GUI ripping: \emph{touch} and \emph{long touch}, which execute a tap and a long tap on a clickable view, respectively; \emph{setText}, which enters a pre-defined text inside an editable view; \emph{keyEvent}, which presses a navigation button; and \emph{scroll}, which scrolls the current window. 

The actual state-based representation of the execution space of an app produced by GUI Ripping consists of the visited states and the executed actions. Test4Enforcers extends the model generated by GUI ripping by adding the information generated by the monitor, that is, the list of the relevant internal events (i.e., the events in the alphabet of the enforcer) executed during each transition. The state-based model thus shows both the UI interactions that can be executed on the app, their effect on the state of the app, and the internal events that are activated when they are executed. 

\begin{figure}[]
\centering
        \includegraphics[width=\textwidth]{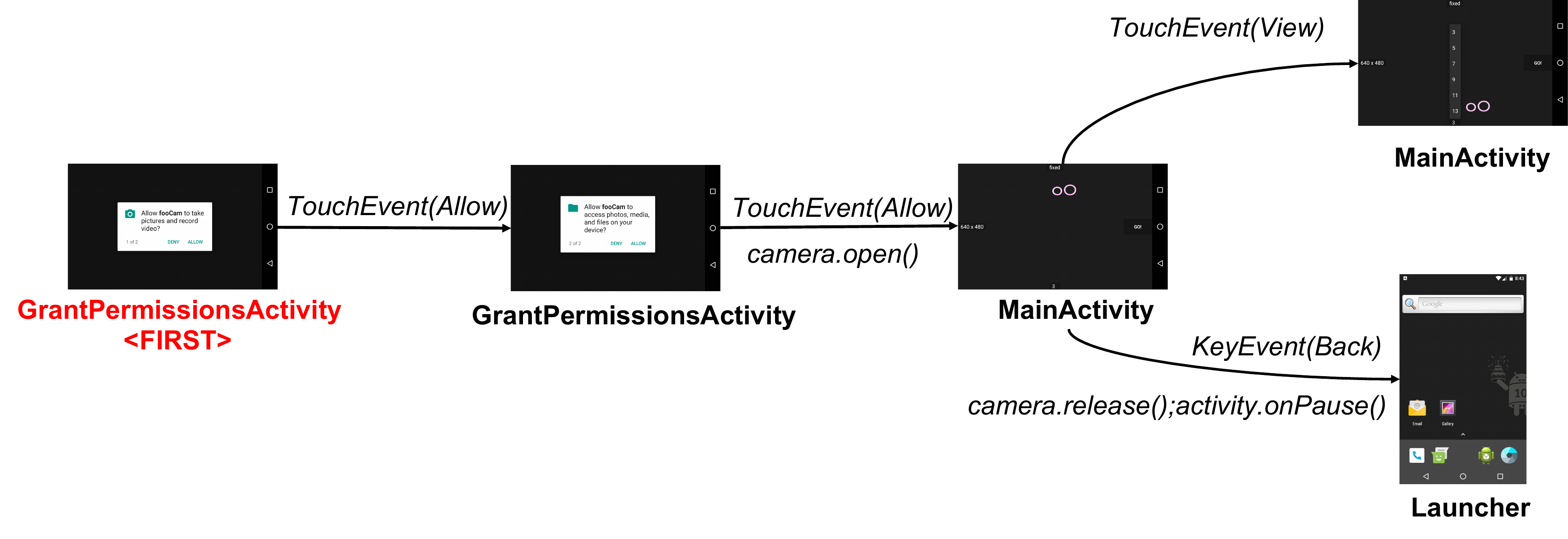}
\caption{Excerpt of a model derived with GUI Ripping.}\label{fig:guimodel}
\end{figure}

More formally, the model resulting from the GUI ripping phase is a tuple $(S_{app}, s_0, T_{app})$, where $S_{app}$ is the set of visited states, $s_0 \in S_{app}$ is the initial state, $T_{app} \subseteq S_{app} \times A_{app} \times in^* \times S_{app}$ is a set of transitions $\langle s_1, a_{app}, \langle in_1, \ldots, in_k \rangle,$ $s_2  \rangle$, where $s_1$ and $s_2$ are the source and target states of the transition, respectively, $a_{app}$ is the UI interaction that causes the transition, and $\langle in_1, \ldots, in_k \rangle$ is a possibly empty sequence of internal events observed during the transition (note these events are exactly the input actions of the enforcer). The resulting model includes everything needed to obtain the concrete test cases (i.e., the sequences of UI operations that must be performed on the app) that cover the test sequences derived with HSI (i.e., the sequences of input operations of the enforcer that must be generated). Figure~\ref{fig:guimodel} shows an excerpt of the model obtained by running the ripping phase on the \texttt{fooCam} app while considering the alphabet of the enforcer shown in Figure~\ref{fig:enforcer-composition}. The  \texttt{fooCam} app is briefly introduced in Section~\ref{sec:caseStudy}. For simplicity, we represent the states with the screenshots of the app. The labels above transitions represent UI interactions, while the labels below transitions, when present, represent internal events collected by the monitor. For instance, when the \emph{KeyEvent(Back)} UI interaction is executed and the app moves from the state \emph{MainActivity} to the state \emph{Launcher}, the sequence of internal events \emph{camera.release()} \emph{activity.onPause()} is observed.  

Note that Test4Enforcers assumes that the software under test has a deterministic behavior, that is, an action performed on a given state always produces the same computation. For Android apps this is often true, and even if this constraint is sometime violated, the test case generation algorithm presented in the next section can compensate this issue by trying multiple sequences of operations, until emitting the correct sequence of events. However, if the behavior of the tested software is highly non-deterministic, it might be difficult for Test4Enforcer to obtain the right set of test cases. 

\subsection{Generation of Concrete Test Cases}
Generating concrete (i.e., executable) test cases consists of finding the sequences of GUI interactions that cause the execution of the desired sequences of internal events, as identified by the HSI method. To this end, Test4Enforcers exploits both the state-based model of the application under test and the sequences generated with the HSI method. Algorithm~\ref{alg:gen} shows the generation process, which takes an app, its GUI model, and a set of sequences to be covered as input, and returns a test suite that contains executable test cases that cover the identified sequences of events. When the algorithm returns the test cases, the mapping between the test case and the covered sequence is maintained. 

Algorithm~\ref{alg:gen} starts by initializing the test suite to the empty set (line~\ref{alg:init}), then for each sequence in the set of sequences to be covered, the algorithm searches for a test that covers the sequence (\texttt{for} loop starting at line~\ref{alg:forLoop}), and, if successful, it both adds an oracle to the test and adds the test to the test suite (lines~\ref{alg:oracle} and~\ref{alg:testsuite} respectively). To identify the concrete test case that can cover a sequence, the algorithm searches for one or more paths in the model that generate the desired sequence of events (line~\ref{alg:searchTest}). For instance, if the sequence to be covered is \textit{camera.open()} \textit{camera.release()} \textit{activity.onPause()} and the GUI model is the one in Figure~\ref{fig:guimodel}, the algorithm can derive the sequence \textit{TouchEvent(v1)} \textit{TouchEvent(v2)} \textit{KeyEvent(BACK)} as the concrete test to execute. In fact, the execution of the identified UI events is expected to produce the desired computation (based on the labels on the transitions). Since an arbitrarily large number of paths covering the desired sequence can be often determined, and not necessarily any path will deterministically produce the desired set of events internal to the app, the algorithm identifies the N (e.g., N=10) shortest paths of the model that cover the sequence (stored in variable \emph{testCases} in the algorithm). Each path of UI events is tested by actually running the application to make sure that both its execution is feasible and it indeed generates the intended sequence of events (\texttt{for} loop  at line~\ref{alg:forLoop2}). Our implementation of Test4Enforcers uses again DroidBot to reproduce a sequence of UI events.

\begin{algorithm}[H]
\DontPrintSemicolon
  
  \KwInput{App app, RippingModel appModel, TestSequences testSequences}
  \KwOutput{Set of $<$TestCase, TestSequences$>$ testSuite}
  
  $testSuite$ $\leftarrow$ $\emptyset$\; \label{alg:init}
  \ForEach{$ts \in testSequences$}{ \label{alg:forLoop}
  $testCases$ $\leftarrow$ generateEventSequences($ts$, $appModel$)\;  \label{alg:searchTest}
  $isCovered$ $\leftarrow$ FALSE\;
  \ForEach{$tc \in testCases$ $\wedge$ $\neg$$isCovered$}{   \label{alg:forLoop2}
  $isCovered$ $\leftarrow$ runTestCase($tc$, $ts$, $app$)\;   
  \If{$isCovered$}{
        tc $\leftarrow$ $addOracle$($tc$, $ts$)\; \label{alg:oracle}
  	$testSuite$.add($tc$, $ts$)\; \label{alg:testsuite}
 } 
 }
 }
  \KwRet testSuite\;
  \caption{Algorithm for generating concrete test cases}\label{alg:gen}
  \end{algorithm}

If the right test is found, the algorithm embeds a \emph{differential oracle} in the test case, before adding it to the test suite. A differential oracle is an oracle that determines the correctness of a test execution by comparing two executions of the same tests on two different programs. In our case, the compared programs are the app \emph{with} and \emph{without} the enforcer deployed. Test4Enforcers can inject two different differential oracles, depending on the characteristics of the sequence of events $in_1, \ldots, in_k$ covered by the test $tc=a_1, \ldots, a_n$ where the oracle must be embedded: the transparent-enforcement oracle and the actual-enforcement oracle. 

\textbf{\emph{Transparent-Enforcement Oracle}}. If the sequence is not the result of any change performed by the enforcer, that is, it covers a path of the enforcement model where the inputs and the outputs are always the same, the test is annotated as a test that must \emph{produce the same result if executed on both the application with and without the enforcer}. More rigorously,  if the output $o_1, \dots, o_m$ is produced by the enforcement model for the inputs $in_1, \ldots, in_k$, this oracle applies when $k=m$ and $in_i=o_i, \forall i=1\ldots k$. The resulting oracle checks the correctness of the execution by first capturing the intermediate states traversed during test execution, as done during the construction of the GUI model, and comparing them when collected from the app with and without the enforcer deployed. More rigorously, if the states $cs_i$ and $cs'_i$ are the states reached after executing the action $a_i$ on the app without and with the enforcer, respectively, the oracle checks if $cs_i=cs'_i, \forall i=1 \ldots n$. If the check fails, the enforcer \emph{is not non-intrusive}, although it was supposed to be, and a failure is reported. For instance, the input sequence \textit{camera.open()} \textit{camera.release()} \textit{activity.onPause()} is not altered by the enforcer in Figure~\ref{fig:enforcer-composition} and thus the transparent-enforcement oracle is used to determine the correctness of the test that covers this sequence, that is, no behavioral difference must be observed when this sequence is executed in both the app without and the app with the enforcer. 

\textbf{\emph{Actual-Enforcement Oracle}}. If the tested sequence corresponds to a path that requires the intervention of the enforcer, the test is annotated as producing an execution that may produce a different outcome when executed on the app with and without the enforcer in place. In such a case, given a sequence of events $in_1, \ldots, in_k$,  $\exists v, s.t.\ in_i = o_i \forall i<v$ and $in_v \not= o_v$. The resulting oracle checks the equality of the states visited by the test case executed on the app with and without the enforcer until the event $in_v$ is produced, and checks for the possible presence of a difference in the following states. More rigorously, if $a_v$ is the GUI action that generates event $in_v$, the actual-enforcement oracle first checks $cs_i=cs'_i, \forall i<v$. If the check fails, the enforcer is unexpectedly intrusive and a failure is reported. For the remaining portion of the execution, it is not possible to know a priori if the activity of the enforcer must result in an effect visible on the GUI. The actual-enforcement oracle thus looks for such a difference, and if the difference is not found, it only issues a warning, suggesting that the enforcer may have failed its activity. Formally, the oracle checks if $\exists p, s.t., cs_p \not= cs'_p$ with $p \geq v$, if it is not the case the warning is issued.  
For instance, the input sequence \textit{camera.open()} \textit{activity.onPause()} causes the intervention of the enforcer shown in Figure~\ref{fig:enforcer-composition}, which outputs an extra event \textit{camera.release()}. The test corresponding to that sequence is thus labeled as \emph{producing the same result until the \textit{activity.onPause()} event, and a potentially different result afterwards}, and the actual-enforcement oracle is embedded in the test.


\section{Case Study} \label{sec:caseStudy}
As initial validation of the approach, we applied Test4Enforcers to the software enforcer that we implemented from the camera release policy shown in Figure~\ref{fig:enforcer-composition} and validated its behavior when injected in the \texttt{fooCam} app, which is a HDR camera app that can take multiple shots with different exposure settings. The app is available on the Google Play Store\footnote{\href{https://play.google.com/store/apps/details?id=net.phunehehe.foocam2\&hl=EN}{https://play.google.com/store/apps/details?id=net.phunehehe.foocam2\&hl=EN}}. We selected this app because it is rather simple, although realistic, it is open source, and we can thus control its execution by manually cross-checking the impact of the enforcer and the behavior of the test cases generated by Test4Enforcers.

To consider both the scenario in which the app violates and does not violate the policy, we produced a faulty version of the app by removing the invocation to the \texttt{release()} operation that the app performs when paused. In the rest of this section, we refer to the correct and faulty apps as \texttt{fooCam}$_c$ and \texttt{fooCam}$_f$, respectively. Moreover, we indicate the apps augmented with the software enforcer for the Camera policy as \texttt{fooCam}$_{c+\textit{enf}}$ and \texttt{fooCam}$_{f+\textit{enf}}$, respectively.

As reported in Section~\ref{sec:hsi}, Test4Enforcers identified the following 5 test sequences that should be covered to test the enforcer:\\


\noindent $activity.onPause()_{req}$, 

\noindent $activity.onPause()_{req}\ activity.onPause()_{req}$,

\noindent $camera.open()_{req}\ activity.onPause()_{req}$, 

\noindent $camera.open()_{req}\ activity.onPause()_{req}\ activity.onPause()_{req}$, 

\noindent $camera.open()_{req}\ camera.release()_{req}\ activity.onPause()_{req}$ \\

Note that, not all these sequences are necessarily feasible. In fact, depending on the specific implementation of the app, some sequences might be impossible to execute. 

To obtain the concrete test cases, we performed steps 2 and 3 of the technique, that is, we configured the Xposed module~\cite{Xposed2020} that we designed for Test4Enforcers to trace the execution of the events in the alphabet of the enforcer and we ran DroidBot on both  \texttt{fooCam}$_c$ and \texttt{fooCam}$_f$ obtaining two Test4Enforcers models. We configured DroidBot to produce 750 UI events, which correspond to about 30mins of computation. We report information about the size of the resulting models in Table~\ref{tab:modelSize}.

\begin{table}[]
\centering
\caption{\label{tab:modelSize}Size of Test4Enforcers models.}
\renewcommand\arraystretch{1.2}
\begin{tabular}{c|c|c}
\multirow{2}{*}{\textbf{App}} & \multicolumn{2}{c}{\textbf{Test4Enforcers Model}} \\ \cline{2-3}
                                         & \textit{\#States}   & \textit{\#Transitions}  \\ \hline
\texttt{fooCam}$_c$                                  & 54                  & 295                                                                                   \\ \hline
\texttt{fooCam}$_f$                                   & 63                  & 276                                                                     \\ \hline
\end{tabular}
\end{table}

The exploration covered a number of states considered the relative simplicity of the app. The difference in the number of states and transitions is due to the randomness of some choices taken during the exploration activity by the tool. Interestingly, the model can now be used to derive the concrete test cases.

The behavior of the app immediately reveals that some sequences cannot be covered in this case. For instance, since \texttt{fooCam} opens the Camera when the \texttt{MainActicity} is started, it is infeasible to execute the \texttt{MainActivity.onPause()} callback without first executing the \texttt{Camera.open()} API call. As a consequence, all the test sequences starting with an invocation to \texttt{MainActivity.onPause()} without a preceding invocation to \texttt{Camera.open()} are infeasible in both \texttt{fooCam}$_c$ and \texttt{fooCam}$_f$. We would like to remark two aspects: (i) this is true for this app, but it is not necessarily true for another app that may open the camera at a different point of the execution, for instance when a button is pressed and not when an activity is started, thus obtaining more feasible test sequences; (ii) the analysis of the model allows Test4Enforcers to not waste time trying to cover sequences that cannot be covered, focusing on the feasible combination of events.

\begin{table}[]
\caption{\label{tab:testcases}Tests automatically generated by Test4Enforcers.}
\centering
\resizebox{1.0\textwidth}{!}{
\renewcommand\arraystretch{1.4}

\begin{tabular}{c l p{8cm}}
\textbf{App} & \textbf{Test Sequences} & \textbf{Coverage}\\ \hline

\multirow{8}{*}{\texttt{fooCam}$_c$}  &     $activity.onPause()_{req}$ & \texttt{infeasible}: $camera.open()_{req}$ must be the first event  \\ \cline{2-3}
& $activity.onPause()_{req}\ activity.onPause()_{req}$ & \texttt{infeasible}: see the above reason \\ \cline{2-3}
& $camera.open()_{req}\ activity.onPause()_{req}$ & \texttt{infeasible}: $camera.release()_{req}$ is invoked from  $activity.onPause()_{req}$, thus it is impossible to have  $activity.onPause()_{req}$ without  $camera.release()_{req}$ \\ \cline{2-3}
& $camera.open()_{req}\ activity.onPause()_{req}\ activity.onPause()_{req}$ & \texttt{infeasible}: see the above reason \\ \cline{2-3}
& $camera.open()_{req}\ camera.release()_{req}\ activity.onPause()_{req}$ & \texttt{feasible}: it is a legal sequence of operations monitored by the enforcer that does not require its intervention. The corresponding test is thus associated with the \emph{transparent-enforcement oracle}. \\

\hline

\multirow{12}{*}{\texttt{fooCam}$_f$}  &     $activity.onPause()_{req}$ & \texttt{infeasible}: $camera.open()_{req}$ must be the first event \\ \cline{2-3}
& $activity.onPause()_{req}\ activity.onPause()_{req}$ & \texttt{infeasible}: see the above reason \\ \cline{2-3}
& $camera.open()_{req}\ activity.onPause()_{req}$ & \texttt{feasible}: it is the sequence that violates the policy and requires the intervention of the enforcer. The corresponding test is thus associated with the \emph{actual-enforcement oracle}. \\ \cline{2-3}
& $camera.open()_{req}\ activity.onPause()_{req}\ activity.onPause()_{req}$ & \texttt{infeasible}: once the activity is paused, it must be resumed to be paused again; resuming the activity causes the execution of $camera.open()_{req}$ in fooCam, thus the sequence is infeasible. \\ \cline{2-3}
& $camera.open()_{req}\ camera.release()_{req}\ activity.onPause()_{req}$ &\texttt{feasible}: it is a legal sequence of operations monitored by the enforcer that does not require its intervention. In the faulty app this sequence can be obtained by interacting with view elements that cause the release of the camera, even if the camere is not automatically released on $activity.onPause()_{req}$. The corresponding test is thus associated with the \emph{transparent-enforcement oracle}.\\ \hline
\end{tabular}
}
\end{table}
  
Table~\ref{tab:testcases} summarizes the results about the feasibility of covering the test sequences in both apps. Column \emph{App} indicates the app that is tested. Column \emph{Test Sequences} indicates the specific sequences of internal events that must be covered. Column \emph{Coverage} reports the outcome obtained using Test4Enforcers in terms of the capability to cover the corresponding sequence. Interestingly, different sequences are feasible in the faulty and correct apps. 

Test4Enforcers derived test cases that cover all the feasible test sequences. The execution of the tests on the apps with and without the enforcer confirmed the correctness of the enforcer for the \texttt{fooCam} app. In particular, the test case derived from the \texttt{fooCam}$_c$ confirmed that both \texttt{fooCam}$_c$ and \texttt{fooCam}$_{c+\textit{enf}}$ behave the same. In fact, the app is correct and the enforcer simply monitored the execution never altering it. The execution of the two test cases derived from \texttt{fooCam}$_f$ on both \texttt{fooCam}$_f$ and \texttt{fooCam}$_{f+\textit{enf}}$ revealed no differences in the behavior of the app. This raised a warning for the test that was expected to activate the intervention of the enforcer. However, the manual inspection of the execution confirmed that a different behavior was observed, since an extra release operation that makes the execution to satisfy the policy is produced when the enforcer is in place. In this specific case, to turn the impact of the enforcer into a visible behavior the test should open a different app that uses the camera, which is outside the capability of DroidBot. 

We can conclude that Test4Enforcers interestingly generated different test cases based on the implementation of the app under test to validate the effect of the enforcers while covering the most relevant feasible test sequences.

\section{Related Work}\label{sec:related}

The contribution described in this paper spans three related research areas: runtime enforcement, model-based testing, and verification of runtime enforcement.

\smallskip
\emph{Runtime enforcement} solutions can be used to prevent a software system from behaving incorrectly with respect to a set of known policies. In particular, runtime enforcement strategies modify executions assuring that policies are satisfied despite the potentially incorrect behavior of the monitored software~\cite{Barringer2010,survey2012}. Enforcement strategies can be specified using a variety of models, including models specifically designed to represent runtime enforcement, such as All or Nothing Automata~\cite{Bielova2011}, Late Automata~\cite{Bielova2011}, Mandatory Results Automata~\cite{Dolzhenko:MRA:2015}, and Edit Automata~\cite{Ligatti:EditAutomata:2005}. 
Runtime enforcement solutions have been applied in multiple application domains, including mobile applications~\cite{Riganelli:HealingDataLos:IWSF:2016,riganelli2017policy,Falcone:AndoridEnforcement:RV:2012,DaianFMSSIR15} operating systems~\cite{Sidiroglou_AAS_2009}, web-based applications~\cite{Magalhaes_SSH_2015}, and cloud systems~\cite{Dai_SHD_2009}.
An overview of techniques  to prevent failures by enforcing the correct behaviour at runtime has been recently published by Falcone et al.~\cite{Falcone2018}. 
Among these many domains, in this paper we focus on the Android environment, which has been already considered in the work by Falcone et al.~\cite{Falcone:AndoridEnforcement:RV:2012}, who studied how to enforce privacy policies by detecting and disabling suspicious method calls, and more recently by Riganelli et al.~\cite{riganelli2017policy,Riganelli:ProactiveLibraries:ACMTAAS:2019,Riganelli:EnforcerReusability:ISOLA:2018}, who studied how to augment classic Android libraries with proactive mechanisms able to automatically suppress and insert API calls to enforce resource usage policies.





While runtime enforcement strategies focus on the definition of models and strategies to specify and implement the enforcers, Test4Enforcers is complemental to this effort, since it derives the test cases that should be executed on applications with and without the enforcers to verify the correctness of the implemented enforcer. 



\smallskip
\emph{Model-based testing} (MBT) refers to the automatic generation of a suite of test cases from models extracted from requirements~\cite{MDT99,DiasSurvey}.
The purpose of the generated test suite is to determine whether an implementation is correct with respect to its specification. MBT approaches are often organized around three main steps~\cite{taxonomy}: building the model, choosing the test selection criteria and building the test case specifications, and generating tests. MBT has been extensively used in the software safety domain, where conformance of the implementation with respect to the model is critical, as shown in the survey by Gurbuz et al.~\cite{GurbuzSurvey}. Test4Enforcers is also a MBT approach, in fact it uses a model, it defines a coverage criterion, and it generates the corresponding test cases.  



A variety of models have been used to guide test case generation, including finite state machines, UML diagrams (statechart, class, activity, and others), and Z specifications~\cite{DiasSurvey}. Indeed, finite-state models are among the most used ones~\cite{7415960}. Interestingly, there are various methods to derive test cases from finite-state machines. For instance, the W~\cite{chow1978testing}, Wp~\cite{Wp}, UIO~\cite{UIO}, DS~\cite{DS}, HSI~\cite{Petrenko1996}, and the H~\cite{HMethod} are well-know test derivation methods~\cite{DOROFEEVA20101286}. Test4Enforcers exploits HSI due to the characteristics of the models used to represent the behavior of the enforcers. Furthermore, Test4Enforcers defines a strategy to produce the target sequences of events while interacting with the UI of an application.

\smallskip
\emph{Verification of runtime enforcement} concerns with checking that the software enforcer is indeed delivering the intended behavior. In fact, although the enforcer is meant to correct the behavior of a monitored software, the enforcer itself might still be wrong and its activity might compromise the correctness of the system rather than improving it.  
A recent work in this direction is the one by Riganelli et al.~\cite{riganelli2017verifying} that provides a way to verify if the activity of multiple enforcers may interfere. The proposed analysis is however entirely based on the models and the many problems that might be introduced by the actual software enforcers cannot be revealed with that approach. Test4Enforcers provides a complemental capability, that is, it can test if the implementation of the enforcer behaves as expected once injected in the target system.


\section{Conclusions} \label{sec:conclusions}
Runtime enforcement is a useful technique that can be used to guarantee that certain correctness policies are satisfied while a running software application. However, specifying enforcement strategies and implementing the corresponding software enforcers might be challenging. In particular, translating an enforcement model into a software enforcer might be difficult because of the significant adaptation and instrumentation effort required to close the gap between the abstraction of the models and the actual implementation, which must take under consideration the requirements of the target execution environment. Indeed, enforcers may easily introduce side effects in the attempt of modifying executions. These are well-known shortcomings of software enforcement solutions~\cite{Bielova2011,riganelli2017verifying}. 

To address these problems, this paper describes Test4Enforcers, a test case generation technique that can automatically derive a test suite that can be used to validate the correctness of the software enforcers derived from enforcement models. The resulting test cases are executed on the same application with and without the enforcer in place. The observed behavioral differences are used to reveal faults and issue warnings. 

Although the approach is not specific to the Android environment, in this paper we focus on the case of software enforcement for Android apps. This domain is particularly relevant because the apps that are downloaded and installed by end-users have been often developed by unknown, and potentially untrusted, parties. Enriching the environment with enforcers can improve multiple aspects, including security and reliability. Studying how to apply Test4Enforcers to other domains is indeed part of our future work.

In addition, Test4Enforcers is designed to reveal misbehaviors that relate to the ordering of events, as represented in the enforcement model. There are of course classes of misbehaviours that go beyond the ones considered in this paper. For instance, timed properties can be used as policies and enforcers that take time information into account can be designed~\cite{Falcone:TimedPropEnf:RV:2019}. Extending the test case generation capabilities of Test4Enforcers to other class of properties is also part of our future work.

%
%
%
 \bibliographystyle{splncs04}
 \bibliography{main}
%





\end{document}